\documentclass[11pt,twoside]{article}
\usepackage{./asp2014}

\resetcounters

\begin{document}

\title{Imaging Molecular Gas at High Redshift}
\author{C.L. Carilli and Y. Shao}
\affil{National Radio Astronomy Observatory, Socorro, NM 87801, U.S.A.; \email{ccarilli@nrao.edu, yshao@nrao.edu}}

\paperauthor{C.L. Carilli}{ccarilli@nrao.edu}{}{National Radio Astronomy Observatory}{}{Socorro}{NM}{87801}{USA}
\paperauthor{Y. Shao}{}{}{National Radio Astronomy Observatory}{}{Socorro}{NM}{87801}{USA}

\begin{abstract}
We perform simulations of the capabilities of the next generation Very
Large Array in the context of imaging low order CO emission from
typical high redshift star forming galaxies at $\sim 1$\,kpc resolution.
We adopt as a spatial and dynamical template the CO 1-0 emission
from M\,51, scaled accordingly for redshift, transition, and
total gas mass.  The molecular gas masses investigated are factors of
1.4, 3.5, and 12.5 larger that of M\,51, at $z = 0.5$, 2, and 4.2,
respectively. The $z = 2$ galaxy gas mass is comparable to the lowest
mass galaxies currently being discovered in the deepest ALMA and NOEMA
cosmological CO line surveys, corresponding to galaxies with star
formation rates $\sim 10$ to 100\,$M_\odot$ yr$^{-1}$.  The ngVLA
will perform quality imaging at 1kpc resolution of the gas
distribution and dynamics over this disk.  We recover the overall
rotation curve, galaxy orientation properties, and molecular ISM
internal velocity dispersion. The model at $z = 4.2$ corresponds to
a massive star forming main sequence disk 
(SFR $\sim 130\,M_\odot$ yr$^{-1}$).  The ngVLA can obtain
1kpc resolution images of such a system in a reasonable integration
time, and recover the basic galaxy orientation parameters, and,
asymptotically, the maximum rotation velocity.  We compare the ngVLA
results with capabilities of ALMA and the Jansky VLA.  ALMA and the VLA can
detect the integrated low order CO emission from these galaxies, but
lack the sensitivity to perform the high resolution imaging to recover
the dynamics at 1kpc scales. To do so would require of order 1000
hrs per galaxy with these current facilities. We investigate a
'minimal' ngVLA configuration, removing the longest baselines and much
of the very compact core, and find good imaging can still be performed
at 1\,kpc resolution.

\end{abstract}

\section{Introduction}

The next generation Very Large Array (ngVLA), is being considered as
a future large radio facility operating in the $1.2-116$\,GHz range.
The current design involves ten times the effective collecting area of
the VLA and ALMA, with more than ten times longer baselines, providing
mas-resolution, plus a dense core on km-scales for high surface
brightness imaging.  The ngVLA opens unique new parameter space in the
imaging of thermal emission from cosmic objects ranging from
protoplanetary disks to distant galaxies, as well as unprecedented
broad band continuum polarimetric imaging of non-thermal processes
\citep{mck16, car15, sel17, bol17}.

\articlefigure{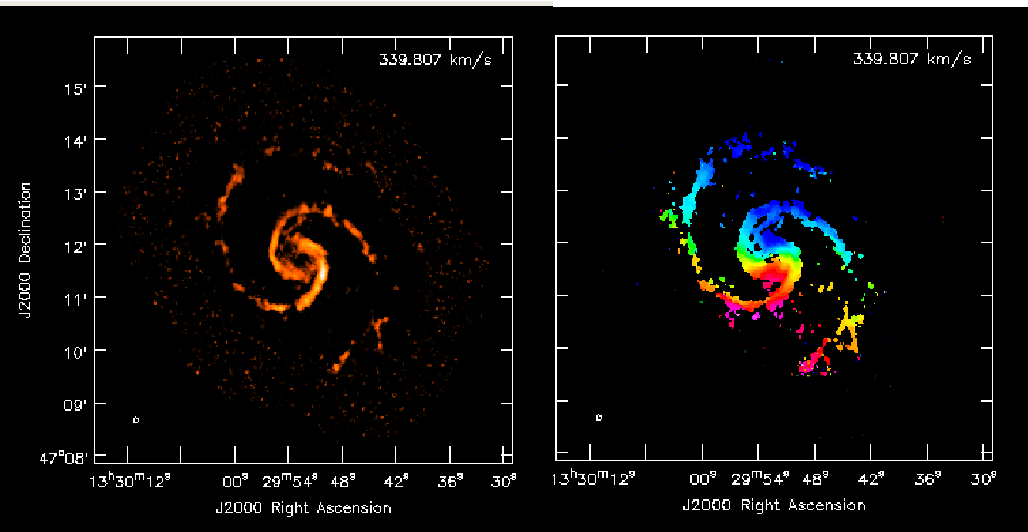}{csfig1}{BIMA SONG CO 1-0 observations of M\,51 at 200\,pc resolution ($5\farcs5$; \citep{hel03}}

One of the primary science drivers for the ngVLA involves study of the
molecular gas content of galaxies throughout cosmic time
\citep{car15}. Cool molecular gas provides the fuel for star formation
in galaxies, and hence represents a key constituent in the study of
the Baryon cycle during galaxy evolution \citep{car13}.  A standard method for
deriving the molecular gas mass in galaxies involves an empirical
calibration of the relationship between the CO 1-0 tracer and total
gas mass \citep{bol13}. The high sensitivity and wide fractional
bandwidth of the ngVLA provides a powerful tool to study the evolution
of the molecular content of galaxies, through observations of the low
order CO transitions.

\begin{figure}[t]
\includegraphics[scale=0.25]{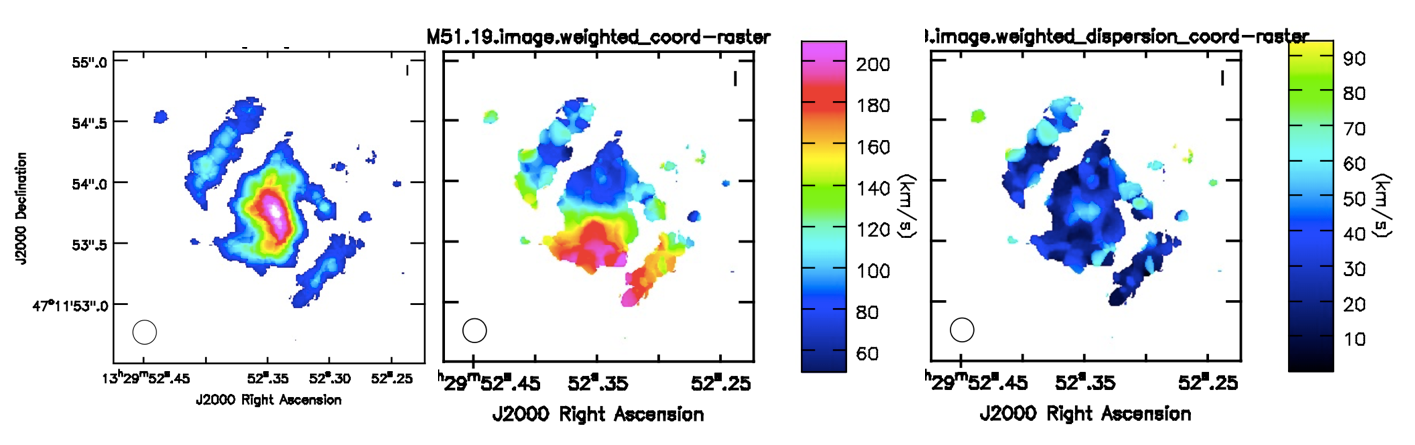}
\includegraphics[scale=0.25]{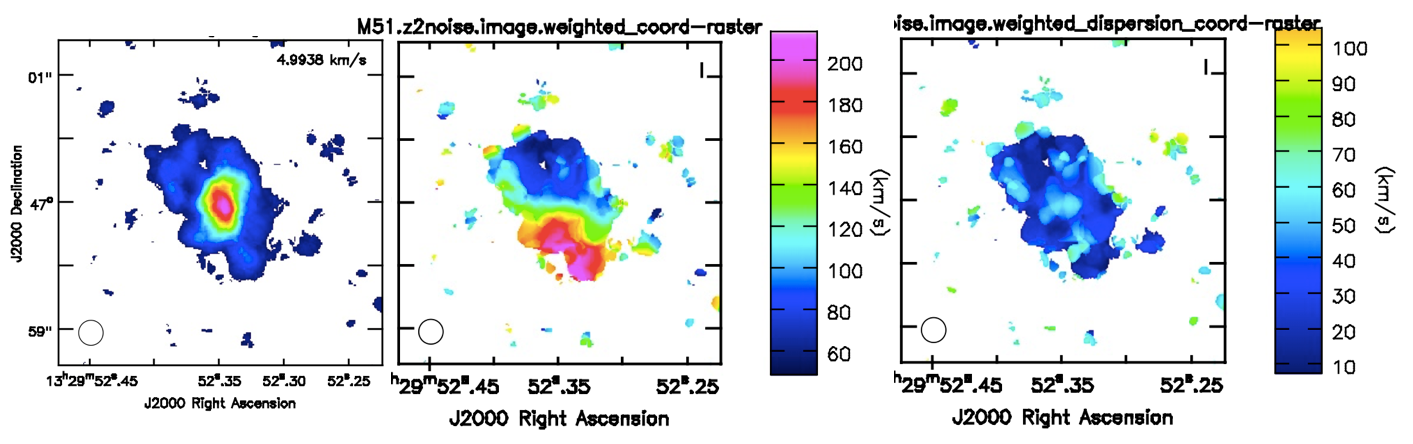}
\includegraphics[scale=0.25]{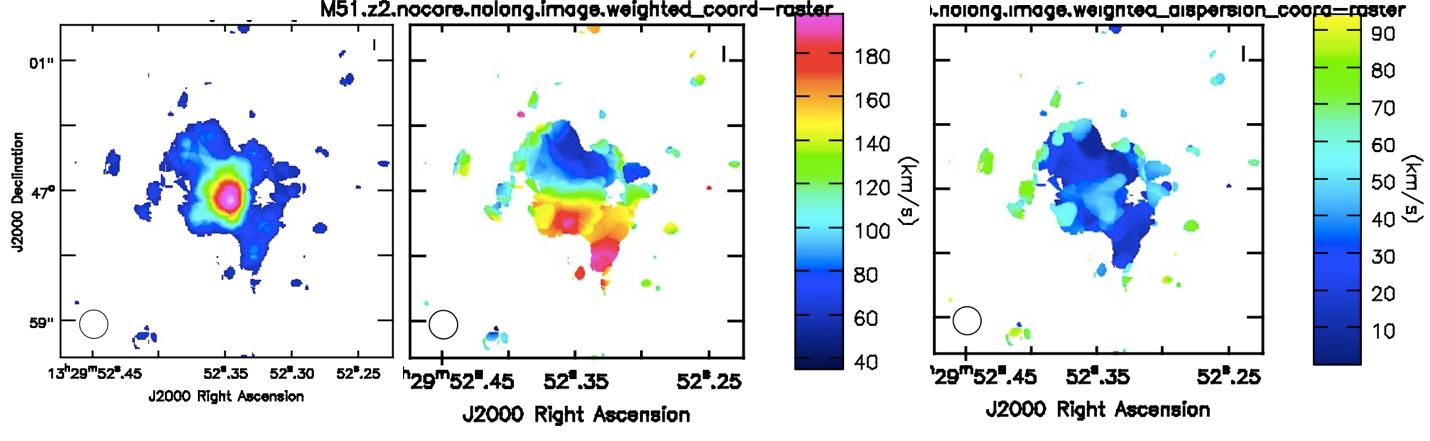}
\includegraphics[scale=0.25]{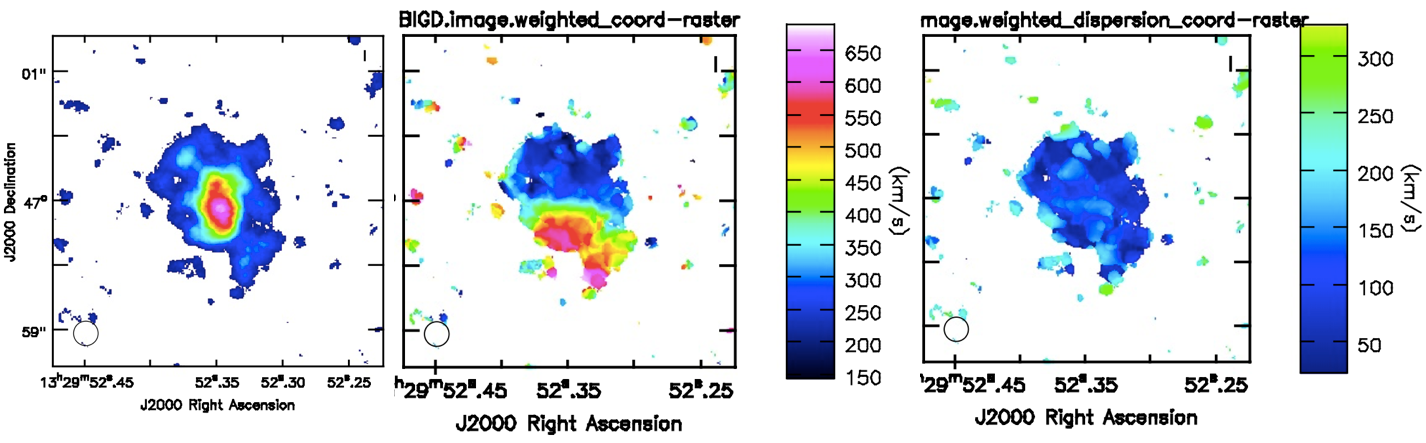}
\caption{({\it Top Row:}) Simulated ngVLA
observations using at 77 GHz of the CO 1-0
emission from an M\,51 analog at $z  = 0.5$ for a 
30\,hr observation ($SFR \sim 7\,M_\odot$ yr$^{-1}$).
The left frame shows
the velocity integrated emission. The middle shows the intensity weighted
mean velocity, and the right shows the velocity dispersion.
(Frames are the same in all four examples).
The spatial resolution is $0\farcs19$. 
({\it Second row:}) Simulated 
ngVLA observation at 77 GHz of the CO 2-1
emission from the $z  = 2$ model galaxy 
(${\rm SFR} \sim 25\,M_\odot$ yr$^{-1}$). 
The spatial resolution is $0\farcs19$.
({\it Third row:}) Simulated observation using a 'minimal 
configuration' of the ngVLA (only 1/3 of the core, and no antennas
at radii outside the original VLA Y), 
at 77 GHz of the CO 2-1 emission from the $z  = 2$ model. 
The spatial resolution is $0\farcs19$.
({\it Bottom row:}) Simulated  ngVLA observation at 
44\,GHz of the CO 2-1
emission from a massive star forming disk galaxy at $z  = 4.2$ 
(${\rm SFR} \sim 130\,M_\odot$ yr$^{-1}$).  
The spatial resolution is $0\farcs22$. 
}
\label{csfig2}
\end{figure}

An area that is currently poorly explored is the detailed distribution
and kinematics of the molecular gas in early galaxies. Progress has
been made using ALMA, the VLA, and NOEMA, but these studies are
time consuming (tens of hrs per galaxy), and remain limited in
spatial resolution (few kpc). The ngVLA will revolutionize this
field, through sub-kpc resolution observations of the molecular
gas down to very low brightnesses in galaxies out to the
highest redshifts \citep{car13}. In particular, the full
frequency range from $30-100$\,GHz, covers both low and high
order CO transitions at high redshift, with 10\,$\times$ the sensitivity
of ALMA and the VLA. 

We investigate the ability of the ngVLA to image the
molecular ISM in high redshift galaxies at $\le 1$\,kpc resolution.
This scale corresponds to the giant molecular star forming clumps
being discovered in active star forming galaxies during the peak epoch
of cosmic star formation, around $z \sim 2$ \citep{gen11}.  We
compare the capabilities of the ngVLA to observe typical `main
sequence' star forming galaxies, with those of ALMA and the VLA. We
also consider a minimal array design with which such measurements
could be made with reasonable accuracy.

\section{Template: CO 1-0 Emission from M\,51}

\articlefigure{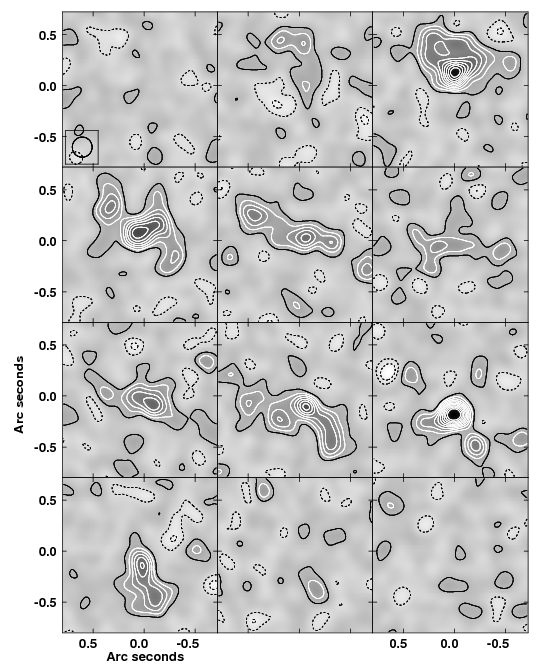}{csfig3}{
Spectral channel images of the CO 2-1 emission from star forming disk galaxy  at $z = 2$ (${\rm SFR} \sim 25\,M_\odot$ yr$^{-1}$), from the 30\,hr ngVLA  observations tapered to $0\farcs19$ resolution.The first contour level is 26\,$\mu$Jy beam$^{-1}$, and the contours are linear. Negative contours are dashed.}


\begin{table}
\footnotesize
\caption{Galaxy Model parameters}
\begin{tabular}{lcccccccc}\hline
Model & CO flux & CO line & $z$ & $d_L$ & $L'_{CO}$ & $M_{H2}$ & $L_{IR}$ & SFR \\
\hline
~ & Jy\,km s$^{-1}$ & ~ & ~ & Mpc & 10$^9$ K\,km/s pc$^2$ & 
($\alpha/3.4$)10$^{10}\,M_\odot$ & $\times 10^{10}\,L_\odot$ & $M_\odot$ yr$^{-1}$ \\
\hline
M\,51 & 10150 & 1-0 & 0.0015 & 7.6 & 1.6 & 0.55 & 4.7 & 4.7 \\
z=0.5 & 0.15 & 1-0 & 0.5 & 2860 & 2.3 & 0.8 & 7.4 & 7.4 \\
z=2 & 0.1 & 2-1 & 2.0 & 15800 & 5.8 & 2.0 & 25 & 25 \\
z=4.2 & 0.1 & 2-1 & 4.2 & 38800 & 20 & 6.9 & 130 & 130 \\
\hline
\vspace{0.1cm}
\end{tabular}
\end{table}

We adopt as a spatial template the CO 1-0 emission from the nearby
star forming disk galaxy, M\,51. M\,51 is one of the best studied galaxies
in molecular gas \citep{sch13}. M\,51 is at a distance of 7.6
Mpc (recessional velocity = 463\,km/s), with an integrated CO 1-0
luminosity of $L'_{CO} = 1.6\times10^{9}$ K\,km/s pc$^2$, implying a
molecular gas mass of $5.5\times 10^9$ ($\alpha/3.4$)\,$M_\odot$, where
$\alpha = 3.4$, corresponding to the Galactic conversion factor for CO
1-0 luminosity to total molecular gas mass. We give the galaxy
properties for M\,51 and the higher redshift models in Table 1.  The
rotation curve of M\,51 is well studied, and the numbers are given in
our comparitive dynamical analysis below.  We employ the public
BIMA SONG CO 1-0 data cubes \citep{hel03}, as the starting
point of the models. The moment 0 and 1 images (velocity-integrated CO
emission and intensity weighted mean velocity) for these data are
shown in Figure 1.

\articlefigure{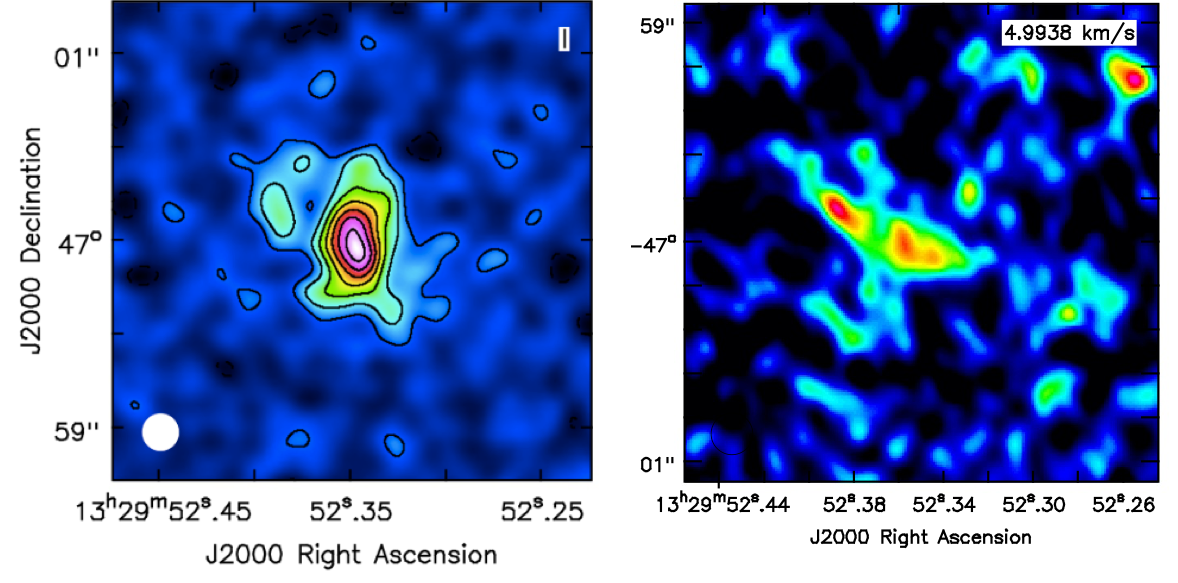}{csfig4}{
({\it Left:}) The ngVLA velocity integrated CO emission as per Figure 2, row 2, for comparison. 
({\it Right:}) Simulated observations using ALMA at 77\,GHz of the velocity-integrated total CO 2-1 emission from the $z  = 2$ model. 
}


We then redshift the source to $z = 0.5$ and $z = 2$, and generate
mock observations of the CO emission with the ngVLA and ALMA.  For the
$z = 0.5$ calculation, we adopt the 1-0 line, redshifted to 77\,GHz. For
the $z = 2$ calculation, we adopt the CO 2-1 line, redshifted to
77\,GHz. We assume thermalized line excitation at least to the 2-1 line,
and hence increase the line flux densities (in Jy), by a factor four
over the 1-0 flux densities.  The CO 1-0 luminosity has been one of
the primary diagnostics for deriving total molecular gas mass in
galaxies \citep{bol13}.  One of the issues in the study of
high redshift galaxies historically is that many observations of CO
emission have employed higher order transitions, thereby requiring an
assumption about the excitation ladder of CO to derive a total
molecular gas mass.  Observing CO 1-0 directly avoids this
extrapolation. Moreover, it has been found that the 2-1 line is almost
always thermally excited (i.e., a luminosity four times that of 1-0, in
Jy\,km s$^{-1}$), in high $z$ galaxies, hence a factor 4 extrapolation
is justified.  However, as shown by Casey et al. (2015), going to
higher order leads to substantial uncertainty in extrapolation to the
1-0 luminosity due to varying sub-thermal excitation in different
galaxies. Hence, the importance of observing the low order transitions
\citep{dad15, car13}.

An important recent discovery with regards to the molecular gas in
galaxies is the observation of a rapid increase in the gas mass to
stellar mass ratio with redshift. Studies of main sequence star
forming disks galaxies show an order of magnitude increase in this
ratio from $z = 0$ to 2. The peak epoch of cosmic star formation
corresponds to an epoch when the baryon content of typical star
forming disk galaxies is dominated by gas, not stars 
\citep{tac10,tac18, gen15, sch16, dad10, gea11, sco17, dec16}. 
To make some
allowance for this increase in gas mass fraction with redshift, we
increase the CO luminosity by a factor 1.4 for the $z = 0.5$ model,
and 3.5 for the $z = 2$ model, relative to M\,51.

\articlefigure{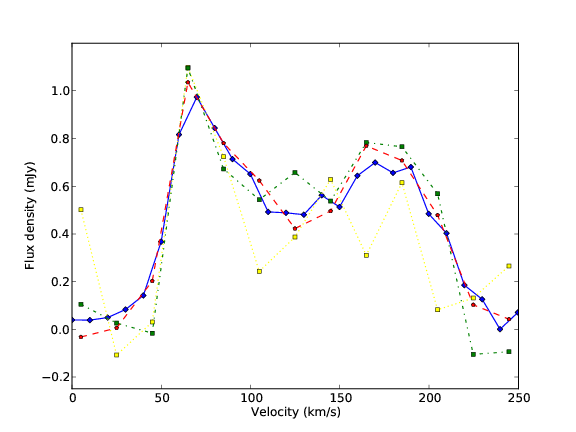}{csfig5}{
The spatially integrated CO 2-1 spectrum of the $z  = 2$
model galaxy. Blue is the input model with no noise added, and
10\,km s$^{-1}$ channel$^{-1}$. The red is for the tapered ngVLA 30hr
observation shown in Figure 2, row 2, at 20\,km s$^{-1}$ channel$^{-1}$.
The yellow is for the ALMA observation shown in Figure 5, at 20\,km s$^{-1}$ channel$^{-1}$.
The green is for the minimal ngVLA array (30\% core, Y-only), 30hr
observation shown in Figure 2, row 3, at 20\,km s$^{-1}$ channel$^{-1}$.}


The CO 2-1 luminosity for the $z = 2$ model is then: $L'_{CO} =
5.8\times10^{9}$ K\,km/s pc$^2$, implying a molecular gas mass of $2.0\times 10^{10}$ ($\alpha/3.4$)\,$M_\odot$.  This is at the extreme
low end of the galaxies currently being discovered in the deepest
ALMA, VLA, and NOEMA cosmological surveys for molecular line emission
from $z \sim 2$ galaxies \citep{wal16,wal14,dec16}. Using
the standard relationship between IR luminosity and CO luminosity for
main sequence galaxies leads to: $L_{IR} \sim 2.5\times 10^{11}\,L_\odot$, and a star formation rate $\sim 25\,M_\odot$ yr$^{-1}$
\citep{dad10}.

\articlefigure{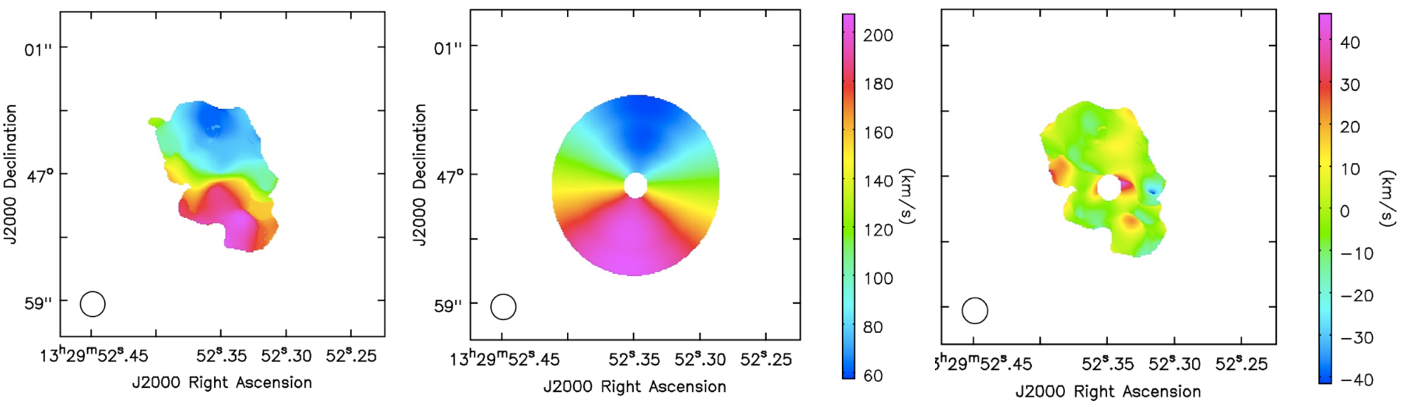}{csfig6}{
({\it Left:}) the mean velocity of the CO emission from the $z=2$ model derived from the ngVLA using the XGAUS technique described in \S5.3.  
({\it Center:}) the model for the rotation curve derived using ROTCUR. 
({\it Right:}) the residuals from differencing the model and observation.
}


We also adopt the CO 2-1 line at $z = 2$ because it will be less affected
by brightness contrast relative to the CMB \citep{zha16}. 
Since all observations are made relative to the mean background of the 
CMB, the increasing temperature of the CMB with redshift can depress the 
observed surface brightness of the CO and dust emission of high redshift
galaxies, with the effect becoming more noticeable for low order
transitions. Their calculation shows that at $z \sim 2$, the effect
on the CO 2-1 brightness should be minimal. 

\articlefiguretwo{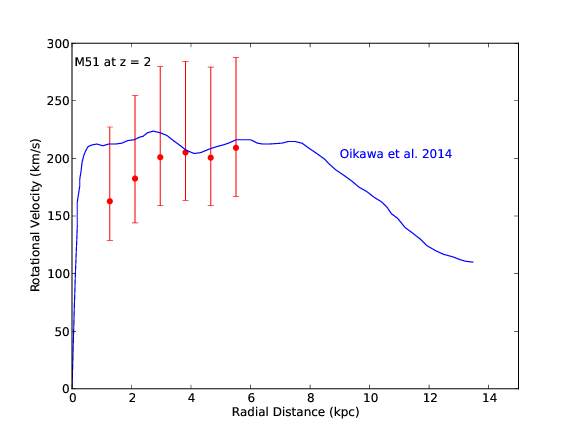}{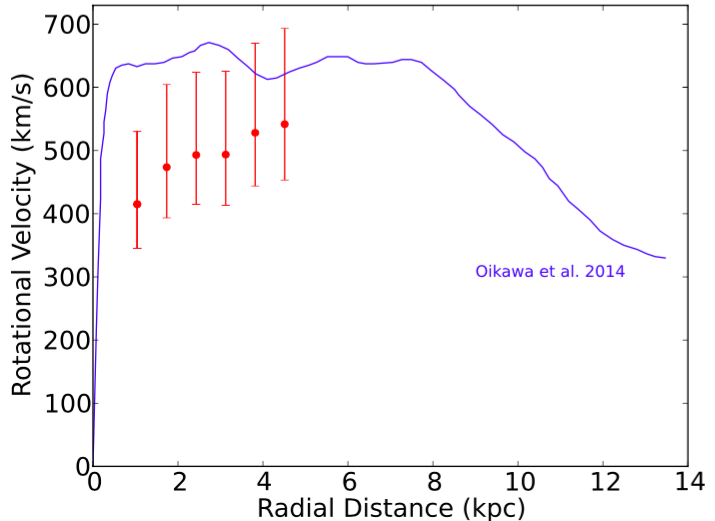}{csfig7}{
({\it Left:}) The rotation curve derived for the $z=2$ model from the ngVLA data using ROTCUR (red points plus errors). The blue line shows the true rotation curve \citep{oik14}.
({\it Right:}) The rotation curve derived for the $z=4.2$ massive star forming disk galaxy, derived from the ngVLA data using ROTCUR (red points plus errors). The blue curve shows the input model scaled by a factor 3 \citep{oik14}.
}


As a final comparison, we consider a galaxy at $z = 4.2$ with
$3.5\times$ higher CO luminosity than the $z=2$ model ($12.5 \times$
M51), and a total velocity width a factor three larger. We adopt the
CO 2-1 line, redshifted to 44\,GHz.  This system would correspond to the
more massive star forming disk galaxies seen at high redshift \citep[SFR $\sim 130\,M_\odot$ yr$^{-1}$;][]{cas14}.

\section{Configuration}


We employ the Southwest configuration, distributed across New Mexico
and Chihuahua\footnote{These
simulations were performed with a slightly older, 300 antenna
configuration, vs. the latest 214 antenna configuration. However, the
($u,v$)-weighting employed down weights both the longer spacings (beyond
30\,km), and the core (within 1\,km), while the intermediate spacing array
is much the same as the latest configuration. Hence, the results will
not change appreciably using the current 214 antenna main array.}. The array includes 45\% of the
antennas in a core of diameter $\sim 1$\,km, centered on the VLA
site. Then some 35\% of the antennas out to VLA A array baselines of
30\,km, and the rest to baselines as long as 500\,km, into Northern
Mexico.

\articlefigure{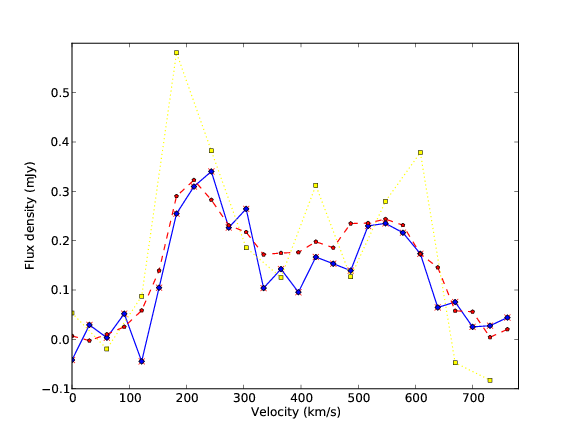}{csfig8}{The integrated spectrum of the CO 2-1 
emission from the $z  = 4.2$
massive star forming disk galaxy. Blue is the input model with no noise and
30\,km s$^{-1}$ channel$^{-1}$. The red is for the tapered ngVLA 30hr
observation shown in Figure 2, row 4, at 30\,km s$^{-1}$ channel$^{-1}$.
The yellow is for a 30hr VLA observation, at 60\,km s$^{-1}$ 
channel$^{-1}$.
}


For the ngVLA noise calculation at 77\,GHz, we assume the radiometer
equation, using an 18m diameter antenna, with 70\% efficiency, 70K
system temperature at 77\,GHz, and a 30\,hr observation.  For the ngVLA
noise calculation at 44\,GHz, we use an 18m diameter antenna, with 75\%
efficiency, 55K system temperature, and a 30\,hr observation.

For the ALMA simulations, we use the ALMA-out18 and ALMA-out22
configurations.  These have 50 antennas extending to baselines that
give a naturally weighted beam at 77\,GHz of $0\farcs45$ and $0\farcs19$,
respectively. For the ALMA noise calculation, we assume antennas of
12m diameter, with an efficiency of 75\%, and 70K system temperature,
and a 30hr integration. Note that these observations require
ALMA band 2, currently under design. 

\articlefigure{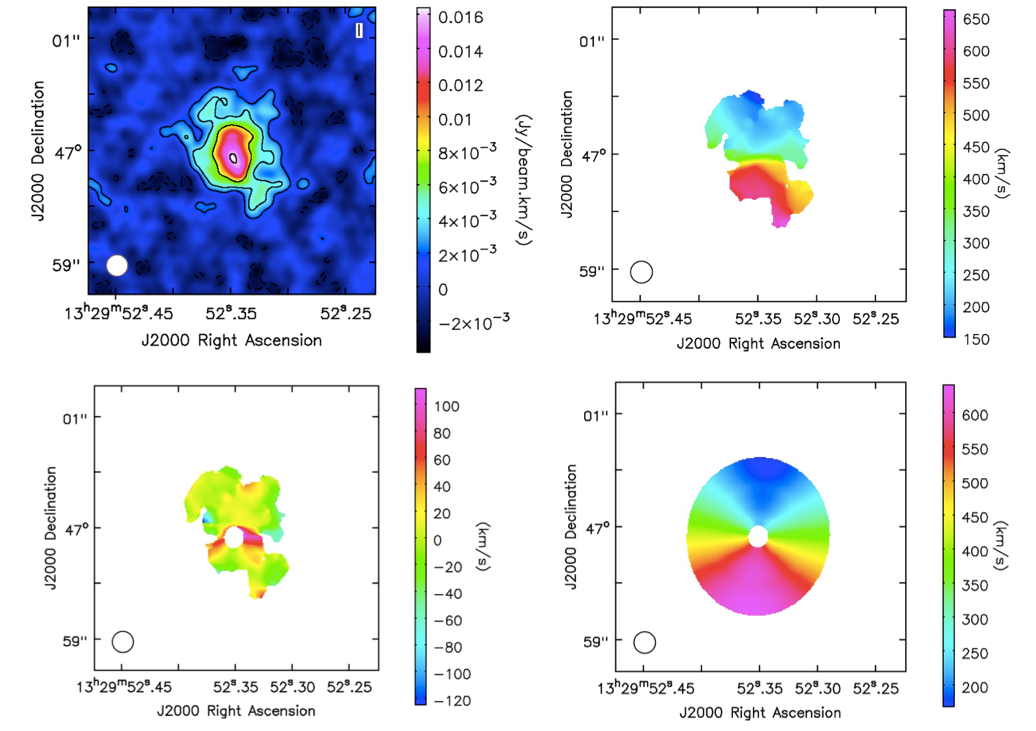}{csfig9}{
({\it Upper Left:}) the velocity integrated CO 2-1 emission from the $z=4.2$ galaxy, 
({\it Upper Right:}) the mean velocity, derived from the ngVLA using the XGAUS technique described in \S5.3.  
({\it Lower Right:}) the model for the rotation curve derived using ROTCUR. 
({\it Lower Left:}) the residuals from the model and observation difference.
}


For the VLA simulation, we employ the B configuration, with 
27 antennas of 25m diameter, with an efficiency of 35\%, and 65K
system temperature, and a 30\,hr observing time.

The noise in the final images is also dictated by the channel width and
visibility weighting, with natural weighting (robust = 2) giving the
optimal noise performance.  The different ALMA configurations are
designed to give a reasonable synthesized beam for natural
weighting. 



The ngVLA has a very non-uniform antenna distribution.  The naturally
weighted beam for this centrally condensed distribution leads to a PSF
with a high resolution core of a few mas width at 77 GHz, plus a
broad, prominent pedestal or plateau in the synthesized beam with a
response of $\sim 50\%$ over $\sim 1\arcsec$ scale.  This prominent pedestal
leads to severe problems when trying to image complex structure. The
imaging process entails a balance between sensitivity and synthesized
beam shape, using visibility tapering and Briggs robust weighting \citep{cjs18, cot17,car17,car16,crbc16,car18}.  

\section{Mechanics of the Simulation}

The starting model is the BIMA SONG observations of CO 1-0 at about
$5\farcs5$ resolution \citep{hel03}.  This corresponds to 0.20 kpc
at the distance of M\,51. The channel width is 10\,km s$^{-1}$ = 3.8\,MHz
at 115GHz. Details of the process can be found in \citet{cs17}.
In brief, we adjust the frequency, pixel size, and flux density of the model
to match the line luminosity and galaxy size as a function of redshift.



We employed the {\sc simobserve} task within CASA to generate uv data sets.
Instructions on how this is done are found on the ngVLA web page. We
simulated a 30\,hr observation, made up of a series of 3\,hr scheduling
blocks around transit. 

\begin{table}[h]
\footnotesize
\caption{Imaging parameters}
\begin{tabular}{lcccccc}\hline
Array & Frequency & Channel Width  & Robust & ($u,v$)-Taper & Resolution & rms noise  \\
\hline
~ & GHz & MHz & ~ & arcsec & arcsec & $\mu$Jy beam$^{-1}$ channel$^{-1}$ \\
\hline
ngVLA CO 1-0 & 77 & 5.1 & 0.5 & 0.15 & 0.22 & 16 \\
ngVLA CO 2-1 & 77 & 5.1 & 0.25 & 0.15 & 0.19 & 16 \\
ALMA-out22 & 77 & 5.1 & 2 & ~ & 0.19 & 75 \\
ALMA-out18 & 77 & 5.1 & 2 & ~ & 0.45 & 75 \\
ngVLA-min & 77 & 5.1 & 1 & 0.15 & 0.21 & 24 \\
ngVLA & 44 & 4.5 & 0.5 & 0.1 & 0.19 & 10 \\
VLA & 44 & 4.5 & 2 & ~ & 0.22 & 80 \\
\hline
\vspace{0.1cm}
\end{tabular}
\end{table}

In all cases we employed the CLEAN algorithm with Briggs weighting,
and adjusted the robust parameter and ($u,v$)-taper to give a reasonable
synthesized beam and noise performance. Our target resolution
was $\sim 0\farcs2$, which corresponds to $\sim 1$\,kpc physical for
$z \ge 0.5$. Synthesized beams for different weighting schemes are
given in Table 2. We use a cell size of $0\farcs01$ throughout. 

\section{Results}

\subsection{$z = 0.5$}



We focus our more detailed analysis on the $z  = 2$ model. However, for
reference, we have performed a mock observation of M\,51 at $z = 0.5$. 
Figure 2, row 1 shows the CO 1-0 moment images for a 30hr observation
for the $z = 0.5$ model with the ngVLA.  The observing frequency is 77
GHz.  Our target resolution is $\sim 0\farcs2 \sim 1$ kpc at $z = 0.5$.
To achieve a well behaved synthesized beam (i.e., no pedestal), but
retain reasonable sensitivity, we employ the imaging parameters listed
in Table 1.  The rms per 20\,km s$^{-1}$ channel, is about 16\,$\mu$Jy
beam$^{-1}$. 

We generate moment images of the CO 1-0 emission blanking at surface
brightnesses below about $1.5\sigma$. The results in Figure 2 show that
the ngVLA will make quality images of the velocity integrated intensity,
as well as the overall velocity field.  The emission is dominated by
the central regions, but emission from the spiral arms remains
clear, as well as the large-scale galaxy velocity structure in the
extended regions.  The minimum measured velocity dispersions are 
20km/s, which is at the limit of the measurements.



\subsection{$z= 2.0$}

We next generate a mock observation of the CO 2-1 emission from the
model galaxy at $z = 2$. Again, 30hrs integration is assumed, and the
observing frequency is 77 GHz. The imaging parameters are
given in Table 2. 
Figure 3 shows the resulting CO 2-1 channel images for the $z = 2$
model. The rms per 20\,km s$^{-1}$ channel (5.1\,MHz), is 16\,$\mu$Jy
beam$^{-1}$. The corresponding $3\sigma$ gas mass limit per channel is
$\sim 2\times10^8$ ($\alpha/3.4$)\,$M_\odot$.

Emission is clearly detected over the full velocity range of the galaxy,
starting with the northern arm at low velocity, through the main 
gas distribution at the galaxy center, to the southern arm at high
velocity. 

Figure 2, row 2 shows the moment images for this system derived from the 
ngVLA observations. While the spiral arms are less prominent, the 
emission is still seen over most of the disk, and the velocity
field is reasonable clear.  

Figure 4 shows the integrated spectrum of the CO emission from the
galaxy. The red curve is derive from the mock ngVLA observation. The
blue curve shows the intrinsic emission derived from a model with no
noise added.  The integrated emission profile is very well recovered,
with a mean flux density of 0.65 mJy beam$^{-1}$, and a total velocity
width of 150\,km s$^{-1}$.  

The CO 2-1 luminosity is $L'_{CO} = 5.8\times10^{9}$ K\,km/s
pc$^2$, implying a molecular gas mass of $2.0\times 10^{10}$
(\rm $\alpha/3.4) M_\odot$.  This corresponds to a CO luminosity a
factor $3.5\times$ larger than M\,51 itself.

We next consider a similar 30hr observation using ALMA. To achieve a
similar spatial resolution of $\sim 1$\,kpc, we employ the 'out22'
configuration and natural weighting of the visibilities. This leads
to an rms noise of 75 $\mu$Jy beam$^{-1}$ per 20\,km s$^{-1}$ channel.
Figure 5 shows the velocity integrated emission from the ALMA observation,
compared to that from the ngVLA observation.  ALMA barely detects the 
velocity integrated emission at this resolution. To recover the
dynamics at the level of detail seen in Figure 2, row 2
would required about 700\,hrs of integration time on ALMA. 

We then use a smaller ALMA array (out18 to achieve $0\farcs45$ resolution), to
avoid over-resolving the emission, and again integrating for 30hrs.
The spatially integrated spectrum is shown as the yellow curve in
Figure 4.  The spatially integrated emission is detected in the
spectrum, although still at relatively low signal-to-noise.

\subsection{Rotation Curve Analysis}

We have analyzed our mock CO 2-1 ngVLA observations of the $z = 2$ disk
using the standard tools for galaxy dynamical analysis available in
AIPS and GIPSY.  The details of the process can be found in 
\citet{jon17a}.

In brief, we generate an optimal moment 1 image (intensity weighted
mean velocity), using the XGAUS routine in AIPS. This routine fits a
Gaussian function in frequency (velocity) at each pixel in the cube
to obtain the mean velocity.

We then run the resulting moment 1 image through the ROTCUR program in
GIPSY. This program assumes the gas has a pure circular rotation in a
gas disk, and fits the velocity field with a series of tilted rings as
a function of radius with a given thickness \citep{rog74}. Each ring
has a number of parameters that are used in the fitting, including
rotation velocity, inclination angle of the normal to the line of
sight, and position angle of the projected major axis of the ring on
the sky.

Figure 6 shows the results, including the moment 1 image from XGAUS,
the resulting fit dynamical disk, and the residuals (fit - observed).
The data are reasonably well fit by the tilted ring model. The fit
position angle of the major axis is $172^{\circ}$, as compared to the `true'
value for M\,51 of $162^{\circ}$ \citep{oik14}. The fit inclination
angle is $21^{\circ}$, as compared to $24^{\circ}$ for the nominal 'true' value of
M\,51 \citep{she07}. Note that M\,51 has a warped disk, but the
warp starts at radius $\sim 7$\,kpc, which is outside our sensitivity area
\citep{oik14}.

The fit rotation velocities (corrected for inclination) for the rings
versus radius are shown in Figure 7a. The blue curve is the rotation
curve of M\,51 from \citep{oik14}.  While the error bars are
substantial, the magnitude in the outer rings matches well the 'true'
rotation velocity of $\sim 200$\,km s$^{-1}$.  This plot also shows the
limitations of such a process: recovering the inner steep rise in the
rotation curve within $\sim 1$ to 2\,kpc radius remains problematic
with data that has only 1kpc resolution.

\subsection{$z  = 4.2$}

As a final example, we present CO 2-1 emission from a massive disk
galaxy at $z = 4.2$.  Note that at this redshift, the combination of
CMB excitation, and the fact that observations always measure the
surface brightness relative to the CMB, may become issues.  At $z \sim
4$, for cold molecular clouds (20K), the `CMB effect' could lower the
observed 2-1 surface brightnesses by up to 60\%, while for warmer gas
(50\,K), the CMB effect is $\le 20\%$ \citep{zha16}. Given the galaxy in
question is a massive star forming galaxy, we assume the latter.

We again assume 30hr integration.  The observing frequency is 44\,GHz.
The imaging parameters are given in Table 1, resulting in 
a PSF with a FWHM = $0.19"$ and no pedestal.  The resulting
rms noise per 30\,km s$^{-1}$ channel, is 10\,$\mu$Jy beam$^{-1}$.

Figure 2, row 4 shows the moment images for this system derived from the 
ngVLA observations. While the spiral arms are less prominent, the 
emission is still seen over most of the disk, including the
north and south arms, and the overall velocity field is clear.  

Figure 8 shows the integrated spectrum of the CO emission from this
galaxy. The red curve is derived from the mock ngVLA observation. The
blue curve shows the intrinsic emission derived from a model with no
noise added.  The integrated emission profile is very well recovered,
with a mean flux density of 0.22 mJy beam$^{-1}$, and a total velocity
width of 450\,km s$^{-1}$.  

The implied CO 2-1 luminosity is: $L'_{CO} = 2.0\times10^{10}$ K\,km/s
pc$^2$, implying a molecular gas mass of $6.9\times 10^{10}$
($\alpha/3.4$)\,$M_\odot$.  The implied star formation
rate for this galaxy based on the standard star formation law for disk
galaxies would be $\sim 130\,M_\odot$ yr$^{-1}$ (Daddi et
al. 2010).  Such a galaxy corresponds to a more massive star forming
disk galaxy ('main sequence'), observed at high redshift (Carilli
\& Walter 2013; Casey et al. 2014).

We next consider a 30hr observation using the current VLA. To
achieve a similar spatial resolution of $\sim 1$\,kpc, we employ the
B-configuration and natural weighting of the visibilities, but include
a taper parameter of $0\farcs15$. This leads to an rms noise of 80 $\mu$Jy
beam$^{-1}$ per 30\,km s$^{-1}$ channel, and a spatial resolution
of $0\farcs22$). The signal is not detectable above the noise in the channel
images. We created a spatially integrated spectrum, as shown as the
yellow curve in Figure 8, smoothed to 60\,km s$^{-1}$ channel$^{-1}$. 
The emission can be seen in this integrated spectrum, but at marginal
signal-to-noise. To reach the required signal-to-noise to perform
the imaging in Figure 2, row 4
would require 2000\,hrs with the current VLA. 

We again perform a rotation curve analysis using XGAUS and ROTCUR
on the ngVLA most observations.
Figure 9 shows the results, including the moment images and the
resulting fit dynamical disk, and the residuals (observed - fit).  The
fit position angle of the major axis is $173^{\circ}$, as compared to the
`true' value for M\,51 of $162^{\circ}$ \citep{oik14}. The fit inclination
angle is $26^{\circ}$, as compared to $24^{\circ}$ for the nominal 'true' value of
M\,51 \citep{she07}.

The fit rotation velocities (corrected for inclination) for the rings
versus radius are shown in Figure 7b. The blue curve is the rotation
curve of M\,51 from \citep{oik14}, scaled up by a factor three.  In this
case, the measured rotation curve appears to asymptote to the maximum
rotation velocity at 6kpc radius. This under-estimation over much of
the radius may result from the marginal measurement of the velocity
field on the southern (fainter) spirals arms of the galaxy. These
results indicate the limitations of this kind of analysis, for a given
signal to noise and spatial resolution.

\subsection{Minimal Array}

We investigate a 'miminal array configuration' that would be adequate
to perform the CO imaging at $z = 2$ considered in \$5.2.

We revised the Southwest configuration as follows. First, we remove
all antennas outside the area covered by the current VLA Y
configuration (all antennas beyond 15km radius of the array
core). These antennas are heavily down-weighted due to the desire to
obtain $\sim 0\farcs2$ resolution, which corresponds to baselines of about
4km at 77\,GHz and 8km at 44\,GHz. This removes about 1/3 of the
antennas. Then we remove 2/3 of the core antennas, or 82 antennas. The
core antennas were also down-weighted by the use of the robust
parameter in the imaging analysis above. This leaves us with 114
antennas (out of the original 300), in an array that has good baseline
coverage on scales out to 30km.

We then perform the same mock observation as above for the $z = 2$
model (30hrs). In this case, we could employ robust = 1, along with
the same cell and taper as before, to achieve similar resolution
($0\farcs21$). The rms noise per 20km s$^{-1}$ channel is now 24\,$\mu$Jy
beam$^{-1}$.  

The resulting moment images are shown in Figure 2, Row 3. The resolution is
slightly lower, but the signal to noise is still adequate to recover
all the main structures of the galaxy as seen with the original SW
array.
The green curve on Figure 4 shows the resulting integrated
spectrum. This matches well the true spectrum shown in the blue.

\section{Conclusions}

We have performed a detailed analysis of the capabilities of the ngVLA
to image at 1kpc resolution the low order CO emission from high
redshift main sequence star forming galaxies. The template for the
galaxy size and dynamics is M\,51.  The model galaxies include: (i)
a galaxy with CO luminosity 40\% higher than M\,51 at $z = 0.5$ (star
formation rate $\sim 7\,M_\odot$ yr$^{-1}$), (ii) a modest star
forming main sequence disk galaxy at $z = 2$ (SFR $\sim 25\,M_\odot$
yr$^{-1}$), and (iii) a massive star forming disk 
at $z = 4.2$ (SFR $\sim 130\,M_\odot$ yr$^{-1}$). The main
results are as follows:

\begin{itemize} 

\item The overall gas distribution and dynamics are
recovered in all cases by the ngVLA.

\item Detailed analysis of the mock rotation curves recover the main
galaxy orientation parameters, such as inclination angle and position
angle of the major axis, to within 10\% in all cases.

\item The rotation velocity versus radius is determined reasonably out
to the maximum radius of the CO disk (6\,kpc) for the $z = 0.5$ and $z =
2$, with the exception of the rapid rise in the rotation curve in the
inner 2\,kpc, which cannot be recovered at 1kpc resolution.

\item The rotation curve analysis for the $z = 4.2$ galaxy asymptotes
to the maximum rotation velocity of the system. This is likely due
to the more limited signal-to-noise on the southern spiral arm emission. 

\item ALMA and the VLA can detect the integrated emission from these
galaxies, but neither can recover the dynamics at high spatial resolution.
To do so would require of order 1000\,hrs per galaxy with these current
facilities. 

\item We explore a `minimal array' to perform this measurement involving 
1/3 of the dense 1km core, and no antennas beyond the current VLA
foot-print, giving an array of 114 antennas to radii of about 15\,km.
We can still recover the source dynamics at $0\farcs22$ resolution,
with a noise figure higher by about 50\% relative to the full array
when suitably tapered and weighted.

\end{itemize}

Observing detailed galaxy dynamics is one of the critical, and yet under 
explored, elements in studies of high
redshift galaxies.  Multiple techniques are now being employed, with
the H$\alpha$ and IFU specroscopy in the near-IR employed on large
telescopes for galaxies at $z \sim 2$ \citep{for18}, and
[C{\sc ii}] 158$\mu$m spectroscopy with ALMA at $z \ge 3$ \citep{jon17b}.
The CO observations play a complementary role, tracing the molecular
gas, as opposed to the ionized or neutral atomic gas.

\acknowledgements The National Radio Astronomy Observatory is a facility of the National Science Foundation operated under cooperative agreement by Associated Universities, Inc.


\end{document}